%% file: main.tex
\documentclass[10pt,conference]{IEEEtran}
\IEEEoverridecommandlockouts

\usepackage{mathptmx} % This is Times font

\usepackage{amsmath,amssymb,amsfonts}
\usepackage{comment}
\usepackage{fancyhdr}
\usepackage[normalem]{ulem}
\usepackage[hyphens]{url}
\usepackage[sort,nocompress]{cite}
\usepackage[final]{microtype}
\usepackage{flushend}
\usepackage[bookmarks=true,breaklinks=true,letterpaper=true,colorlinks,citecolor=blue,linkcolor=blue,urlcolor=blue]{hyperref}

\usepackage[utf8]{inputenc}
\usepackage{tikz}

 \usepackage{subfig}
\usepackage{subfig}
\usepackage{array,multirow,graphicx}
\usepackage[linesnumbered,ruled,vlined]{algorithm2e}

\newcommand{\microsubmissionnumber}{XXX}
\fancypagestyle{firstpage}{
  \fancyhf{}
  
  \fancyhead[C]{\vspace{10pt}\normalsize{MICRO 2023 Submission
      \textbf{\#\microsubmissionnumber} -- Confidential Draft -- Do NOT Distribute!!}\\\vspace{-25pt}} 
  \fancyfoot[C]{\thepage}
}

\newcommand{\squishlist}{
	\begin{list}{$\bullet$}
		{ \setlength{\itemsep}{0pt}      \setlength{\parsep}{0pt}
			\setlength{\topsep}{0.5pt}       \setlength{\partopsep}{0pt}
			\setlength{\listparindent}{-2pt}
			\setlength{\itemindent}{-5pt}
			\setlength{\leftmargin}{1em} \setlength{\labelwidth}{0em}
			\setlength{\labelsep}{0.5em} } }
	
	\newcommand{\squishend}{
\end{list}  }

\makeatletter % changes the catcode of @ to 11
\newcommand{\linebreakand}{%
  \end{@IEEEauthorhalign}
  \hfill\mbox{}\par
  \mbox{}\hfill\begin{@IEEEauthorhalign}
}
\makeatother % changes the catcode of @ back to 12

\newcommand{\dashlist}{
	\begin{list}{--}
		{ \setlength{\itemsep}{0pt}      \setlength{\parsep}{0pt}
			\setlength{\topsep}{0.5pt}       \setlength{\partopsep}{0pt}
			\setlength{\listparindent}{-2pt}
			\setlength{\itemindent}{-5pt}
			\setlength{\leftmargin}{1em} \setlength{\labelwidth}{0em}
			\setlength{\labelsep}{0.5em} } }
	
	\newcommand{\listend}{
\end{list}  }

\newcommand{\ceil}[1]{\left\lceil #1 \right\rceil}

\AtBeginDocument{%
  \providecommand\BibTeX{{%
    \normalfont B\kern-0.5em{\scshape i\kern-0.25em b}\kern-0.8em\TeX}}}

\newcommand{\todo}[1]{\textcolor{red}{#1}}

\begin{document}
%-------------------------------------------------------------------------------
%%%%%% -- Title and authors name start-- %%%%%%%%

%don't want date printed
\date{}

\title{\Large \bf DC: Depth Control on Quantum Classical Circuit }

\author{
\IEEEauthorblockN{1st Movahhed Sadeghi}
\IEEEauthorblockA{Pennsylvania State University\\CSE Conference\\mus883@psu.edu}  
%\IEEEauthorrefmark{1}corresponding Author\\

\and
\IEEEauthorblockN{2nd Soheil Khadirsharbiyani}
\IEEEauthorblockA{Pennsylvania State University\\CSE Conference\\szk921@psu.edu}   
\and
\IEEEauthorblockN{3rd Mostafa Eghbali Zarch}
\IEEEauthorblockA{NC State University\\ECE Conference\\ meghbal@ncsu.edu}
\and
\linebreakand
\IEEEauthorblockN{4th Mahmut Taylan Kandemir}
\IEEEauthorblockA{Pennsylvania State University\\CSE Conference\\mtk2@psu.edu}  
}

\maketitle

%%%%%% -- Title and authors name end-- %%%%%%%%

%%%%%% -- PAPER CONTENT STARTS-- %%%%%%%%

\input{sections/0-abstract}

\pagestyle{plain} % removes running headers

\input{sections/1-introduction}

\input{sections/2-background}

\input{sections/3-related}

\input{sections/4-motivation}

\input{sections/5-design}

\input{sections/6-evaluation}
\input{sections/7-conclusion}

%\clearpage 
%%%%%%% -- PAPER CONTENT ENDS -- %%%%%%%%

%%%%%%%%% -- BIB STYLE AND FILE -- %%%%%%%%
\bibliographystyle{unsrt}
\bibliography{refs}
%%%%%%%%%%%%%%%%%%%%%%%%%%%%%%%%%%%%

%%%%%%%%%%%%%%%%%%%%%%%%%%%%%%%%%%%%%%%%%%%%%%%%%%%%%%%%%%%%%%%%%%%%%%%%%%%%%%%%
\end{document}

%% file: sections/0-abstract.tex
%%%%%

\begin{abstract}
The growing prevalence of near-term intermediate-scale quantum (NISQ) systems has brought forth a heightened focus on the issue of circuit reliability. Several quantum computing activities, such as circuit design and multi-qubit mapping, are focused on enhancing reliability via the use of different optimization techniques. The optimization of quantum classical circuits has been the subject of substantial research, with a focus on techniques such as ancilla-qubit reuse and tactics aimed at minimizing circuit size and depth. Nevertheless, the reliability of bigger and more complex circuits remains a difficulty due to potential failures or the need for time-consuming compilation processes, despite the use of modern optimization strategies.

This study presents a revolutionary Depth Control (DC) methodology that involves slicing and lowering the depth of conventional circuits. This strategy aims to improve the reliability and decrease the mapping costs associated with quantum hardware. DC provides reliable outcomes for circuits of indefinite size on any Noisy Intermediate-Scale Quantum (NISQ) system. The experimental findings demonstrate that the use of DC leads to a substantial improvement in the Probability of Success Threshold (PST), with an average increase of 11x compared to non-DC baselines. Furthermore, DC exhibits a notable superiority over the next best outcome by ensuring accurate outputs with a considerable margin. In addition, the utilization of Design Compiler (DC) enables the execution of mapping and routing optimizations inside a polynomial-time complexity, which represents an advancement compared to previously suggested methods that need exponential time.

%The Depth Control (DC) approach, proposed and evaluated in this paper, aims to address this pressing issue by {\em slicing} a given classical circuit into {\em blocks} and {\em reducing} its depth, thereby delivering reliable output and decreased mapping costs for real quantum hardware. Additionally, DC can provide reliable output for infinitely large quantum classical circuits on any NISQ system. Our experiments with DC reveal that it successfully provides reliable output for quantum classical circuits which have previously exhibited little to no output reliability (even after implementing all currently feasible enhancements). Specifically, DC guarantees the attainment of the expected output with a PST higher than 50\% for circuits that do not yield any reliable output when employing the non-sliced baseline. Our results also show that DC achieves, on average, an 11x PST improvement compared to the non-sliced baseline, while ensuring that the correct output is achieved with a substantial gap relative to the second-most reported result. Furthermore, by minimizing the number of gates per job, under DC, all mapping/routing optimizations can be executed in polynomial time, as shown by our theoretical analysis. This achievement was previously unattainable with earlier mapping/routing techniques, which required exponential time. 
\end{abstract}

%%%%%

%% file: sections/1-introduction.tex
%%%%% 
\section{Introduction}
Over the past few decades, quantum computing has witnessed significant advancements, driven by factors such as superior speedup compared to classical systems in select algorithms, as demonstrated   in~\cite{shor1999polynomial,grover1996fast,farhi2014quantum,kandala2017hardware,biamonte2017quantum}, as well as the increased availability of quantum hardware, as evidenced by commercial quantum products. In fact, major vendors, including Google, IBM and Amazon, have developed quantum machines to exploit the promising potential of quantum algorithms. However, due to limitations such as restricted number of qubits and significant errors from various sources, quantum systems are currently unable to reach their full-potential. Furthermore, the existing quantum error correction (QEC) methods, such as~\cite{fowler2012surface,hu2019quantum,campagne2020quantum,ma2020error,google2023suppressing}, have been unable to address these errors in practical systems, due to the excessive number of qubits required to implement them or the excessive number of errors on larger circuits.

To tackle this issue, {\bf Noisy Intermediate-Scale Quantum} (NISQ) systems have been introduced, aiming to execute small-to-medium circuits on quantum  machines, {\em without} relying on error correction techniques. Although this approach offers researchers a way to optimize quantum programs, error rates remain high even for such circuits. Motivated by this, recent works have focused on enhancing reliability by minimizing the impact of errors through mapping~\cite{Liu2022not,zhang2020depth,li2019tackling,tannu2019ensemble,liu2021qucloud,dou2020new}, scheduling~\cite{murali2020software}, routing~\cite{tannu2019ensemble,liu2021qucloud,li2019tackling}, and other optimizations. These techniques have shown relatively reliable outputs when circuit depth is not very high.
%Please note that analogous approaches can be applied to general quantum circuits in the future, when quantum memory technology becomes commercially available. 

Quantum computation is capable of performing all 'classical computations'  while also providing enhanced algorithmic performance opportunities~\cite{shor1999polynomial,grover1996fast,farhi2014quantum,kandala2017hardware,biamonte2017quantum}. Since classical computation has well-known algorithms and is important for the sake of backward compatibility, it is desirable to implement classical logic and computations on quantum computers. This leads to the development of {\bf Quantum Classical Circuits} \cite{parent2015reversible,Paler2016Resizing,ding2020square}, which are basically {\em classical circuits implemented using quantum gates on qubits}. Reversibility is an additional advantage of these circuits over their classical counterparts. For example, the input of a quantum classical AND gate can be uniquely specified from its output, whereas the conventional classical AND gate does not offer this feature (i.e., it is impossible, in the classical domain, to uniquely determine the inputs of an AND gate whose output is 0). This has led to significant interest in optimizing these circuits aggressively~\cite{parent2015reversible,ding2020square,Paler2016Resizing}. Other use-cases for these gates include sub-gates in general quantum algorithms/circuits~\cite{shor1999polynomial,grover1996fast}, as well as research in analyzing classical logic under quantum  superposition state initialization~\cite{ding2020square}.

%It is important to note that optimizations targeting these circuits are treated as general quantum circuit optimizations rather than specifically targeting quantum classical circuits.

While small quantum classical circuits yield reasonably accurate results, reliability decreases as the number of gates and the depth increase. Motivated by this observation, in this paper, we propose and experimentally evaluate a novel approach to address these challenges, enabling the execution of 'infinitely large' quantum classical circuits with 'improved reliability' on NISQ systems. Our motivation arises from the observation that, as quantum systems scale, the reliability of the outputs generated by large quantum circuits declines substantially due to various errors, such as gate errors and coherence errors. Drawing inspiration from classical computing, where dynamic random-access memory (DRAM) cells use 'retention time' based  mechanisms to retain stored data and avoid leakage errors, we attempt to {\em slice} a given large circuit and obtain highly reliable results by manipulating {\em classical memory} for each 'block' and preventing the errors from continuing from one block to the next. This approach can be advantageous if we can ensure the reliability of each block's results, leading us essentially to develop a simple, effective, and universally applicable 'noise filtering mechanism' for quantum classical circuits, enabling in a sense infinite growth in circuit gate counts or depth, while  maintaining output reliability.

In this paper, we present a pioneering approach to address the challenge of error accumulation in quantum circuits, which often results in  unreliable outcome/result. Our method, called {\bf Depth Control}  (DC), focuses on quantum classical circuits; divides such circuits into 'blocks'; and corrects the states of qubits at the end of each block, eventually preventing the errors from propagating over the circuit.  Specifically, we introduce two strategies for this purpose: {\em Static-Depth-Control} (SDC) and {\em Dynamic-Depth-Control} (DDC), both designed to ensure a high probability of obtaining accurate outputs while offering different tradeoffs in terms of reliability and job execution time. Additionally, we tackle the issue of 'reversibility' in quantum circuits after applying our DC approach. We demonstrate that a certain level of reversibility can be maintained, which is important as it can aid in error correction and reliability checks. Furthermore we developed an strategy to apply DC for classical circuit under superposition state inputs. Hence, our proposed DC provides a promising solution to enhance the reliability and performance of quantum classical circuits in the era of NISQ systems. In summary, this paper makes the following key {\bf contributions}:

\squishlist
    \item We introduce, implement, and evaluate DC (Depth Control), an approach that can effectively limit errors ($\epsilon_n$) in quantum classical circuits and enable the execution of infinitely-large circuits.
    \item We develop two different implementation strategies for achieving DC, namely, Static Depth Control (SDS) and Dynamic Depth Control (DDC).
    \item We demonstrate that DC not only ensures the reliability of the output but also produces super-linear speedups for mapping, routing, and other transpile optimizations. In general, our SDC approach achieves a speedup from $\mathcal{O}(m!n!)$ to $\mathcal{O}(n)$, on the mapping/routing algorithm. % ~\footnote{In Qiskit~\cite{Qiskit}, which is an open-source software development kit for working with quantum computers at the level of pulses, circuits and algorithms,  'transpiling' is the process of converting a high-level quantum circuit into a low-level circuit optimized for a particular quantum hardware architecture.}
    \item Additionally, we introduce a general 'uncomputation' strategy for quantum classical circuits {\em after} applying DC. Specifically, by adding a 'reverse DC algorithm', we demonstrate that, despite using a measurement gate in the middle of the circuit, we are still able to preserve circuit's reversibility for classical inputs.
    \item We also develop a novel method for applying DC to quantum classical circuits with the superposition state of inputs. Input superposition states haven't been pertinent to earlier reliability enhancement strategies for quantum classical circuits, such as those mentioned in \cite{ding2020square,parent2015reversible}, due to the constraints imposed by the no-cloning theorem \cite{wootters1982single}. Similarly, they were not relevant to the approach outlined in \cite{Paler2016Resizing}, which will not produce correct result due to quantum measurement teleportation.
    \item Consistent with our theoretical analysis, DC effectively delivers reliable output on real quantum hardware for circuits that initially exhibit little to no reliability. More specifically, our results reveal an 11x PST improvement compared to 'non-blocked' execution, while ensuring the correct output is achieved with a substantial gap relative to the second most reported result -- a feat that the baseline could not achieve.
\squishend

%%%%%

%% file: sections/2-background.tex
%%%%

\section{Background }
\label{sec:Background}
This section first provides an overview of quantum computing, quantum systems, quantum errors, and methods for mitigating them. Next, it discusses quantum classical circuits and their construction. 

\subsection{Quantum Computation and NISQ}
\label{subsec:NISQ} 
Quantum computing operates based on 'qubits', as opposed to classical 'bits'. In classical computation, bits are represented by a value of $0$ or $1$, whereas qubits are represented by a vector specified as $|\phi\rangle=\alpha|0\rangle+\beta|1\rangle$. Upon measurement, a qubit will yield $0$ or $1$ with probabilities $\alpha^2$ and $\beta^2$, respectively, given the constraint that $\alpha^2+\beta^2=1$. When a new qubit is added, the state space expands exponentially (e.g., a two-qubit system will be $\alpha_{00}|00\rangle+\alpha_{01}|01\rangle+\alpha_{10}|10\rangle+\alpha_{11}|11\rangle$), which can be advantageous for certain applications or algorithms. %Indeed, the potential for quantum computing presents fascinating possibilities, with various performance benefits compared to their classical counterparts. In some cases, the magnitude of these improvements can reach exponential levels  compared to the classical algorithms, as demonstrated by, for example, Shor's algorithm~\cite{shor1999polynomial}.

% Numerous algorithms have been introduced over the past few decades to demonstrate the enhancements offered by quantum computing. For instance, Shor's algorithm~\cite{shor1999polynomial} was developed to shift the order of discovering prime factors from classical 'exponential time' to quantum 'polynomial time', whereas Grover's search algorithm~\cite{grover1996fast} provides a square-root improvement compared to its classical counterparts. %Other significant quantum algorithms include Quantum Approximate Optimization Algorithm (QAOA)~\cite{farhi2014quantum}, methods for identifying potential chemical links~\cite{kandala2017hardware}, and machine learning (ML) algorithms~\cite{biamonte2017quantum}.

\begin{comment}
\subsection{Noisy Intermediate Scale Quantum Computation Systems  (NISQ)}
\label{subsec:NISQ} 

\begin{figure}[t!]
\center
% \vspace{-4mm}
\includegraphics[width=0.7\columnwidth]{Figures/NISQSWAP.png}
\caption{(a) SWAP gate implementation, (b) Architecture of the system, (c) Pre-mapping circuit, and (d) Post mapping circuit.}
 \vspace{-6mm}
\label{Fig:NISQSWAP}
\vspace{-0.1in} 
\end{figure} 

\end{comment}

% Due to the potential benefits offered by the aforementioned quantum algorithms,
Several vendors, including IBM, Google, and Intel, are developing quantum hardware solutions. However, to our knowledge, all existing qubit implementation technologies available today are susceptible to numerous errors.
% and current quantum error correction (QEC) codes require, unfortunately, an impractical number of 'parity' qubits, leading to the emergence of the NISQ (Noisy Intermediate Scale Quantum) era.
NISQ systems -Which are the current available hardware- are based on two fundamental principles: i) any  quantum logic can be achieved by combining unitary gates with CNOT gates, and ii)  a system with a limited number of links, multi-qubit operations can still be executed via SWAP gates. Quantum operations supported by quantum machines can be divided into two broad groups. The first group includes unitary gates, which modify either the 'phase', 'value', or both phase and value of a qubit, effectively altering the 'state' of a qubit. Examples of such gates are X gate, Z gate, Y gate, and rotating gates. The second group consists of multi-qubit gates, e.g., the CNOT gate, which is used to {\em entangle} two qubits, causing them to establish a so called 'shared' state. By utilizing three CNOT gates, a SWAP gate can be constructed, which can be used to swap the states of two qubits.

%To illustrate how a NISQ system operates, Figure~\ref{Fig:NISQSWAP}(a) presents a SWAP gate implementation, and Figure~\ref{Fig:NISQSWAP}(b) demonstrates the execution of a sample quantum circuit on a basic NISQ system. Using the mechanism depicted in Figure~\ref{Fig:NISQSWAP}, a NISQ system can execute {\em any} quantum logic as long as it encompasses a full set of unitary operations along with a CNOT, irrespective of the number of connections it possesses. It is important to note that NISQ systems are designed and optimized for small-to-medium sized quantum circuits, and due to their error-prone and noisy nature, they inevitably {\em fail} when targeting large circuits.   

\subsection{Quantum Errors}
\label{subsec:QuantumErrors}
Quantum errors in a NISQ system arise from the imperfect physical nature of the underlying system. There are three fundamental types of quantum errors that can occur during the execution of a quantum circuit:

\squishlist
    \item {\em Gate Error:} This error results from the imperfect implementation of quantum gates and system noise. Since a 'measurement gate' is also a quantum gate, it can be included in this error type.
    \item {\em Coherence Error:} When a qubit remains idle, its state approaches $|0\rangle$ exponentially over time with respect to two constant terms --  $T_1$ and $T_2$. $T_1$ is the constant reflecting the transition from state $|1\rangle$ to state $|0\rangle$, and $T_2$ represents the transition from state $|+\rangle$ to state $|0\rangle$. The larger $T_1$/$T_2$, the smaller the exponent, and consequently, the lower the coherence error. 
    \item {\em Crosstalk Error:} When two gate operations occur simultaneously on adjacent qubits, there is a probability that their states would {\em flip}. This can be more problematic when both operations involve 'entangled' gates, such as a CNOT gate. 
\squishend 

These errors are one of the primary concerns when executing quantum circuits on real quantum hardware. Various mapping, circuit design, gate scheduling, and QEC strategies (e.g., see \cite{fowler2012surface,hu2019quantum,campagne2020quantum,ma2020error,google2023suppressing} and the references therein) have been developed to mitigate these errors and achieve reliable output from a quantum circuit.

\subsection{Quantum Classical Circuits}
\label{subsec:QuantumClassicalCircuits} 
\begin{figure}[h]
\center
\vspace{-3mm}
\includegraphics[width=0.9\columnwidth]{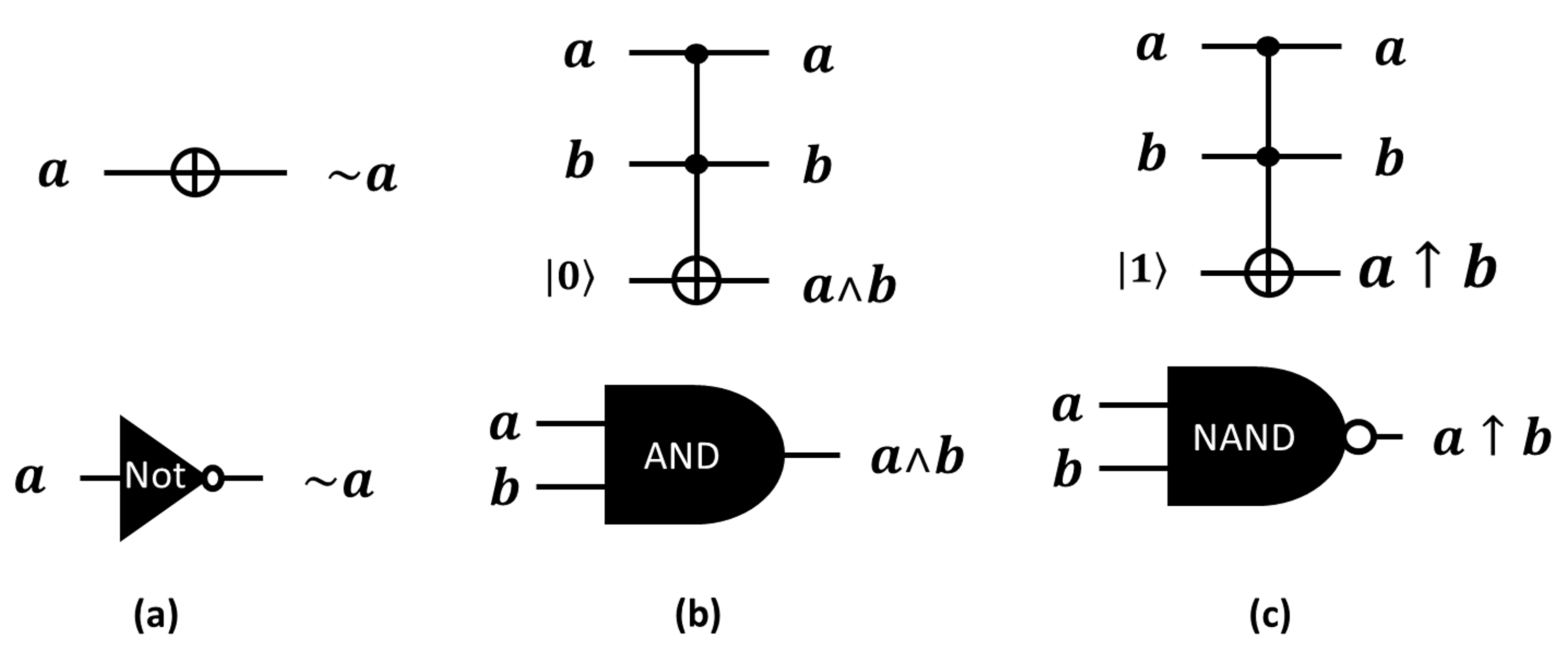}
\caption{Implementing (a) Not gate; (b) 2-input AND gate; and  (c) 2-input NAND gate.}
% \vspace{-6mm}
\label{Fig:AND}
\vspace{-0.1in} 
\end{figure}

Quantum classical circuits are a collection of classical circuits implemented using quantum operations and qubits. The primary reason for implementing quantum classical circuits is to take advantage of the additional computational capacity provided by quantum computing. It is important to note that {\em all} classical circuits can be implemented using quantum operations. To implement a classical circuit on a quantum machine, we need to identify how a NAND gate can be implemented using quantum operations, as it is well-known that the NAND gate is sufficient to construct any classical logic. To demonstrate this, a NAND gate is first divided into an AND gate followed by a NOT gate. Then, by employing MCT (Multi-Controlled Toffoli) circuits, we can implement all these circuits, as illustrated in Figure~\ref{Fig:AND}.

The AND operation yields a true output when both inputs are 1 and a false output otherwise. The Toffoli gate, as depicted in Figure~\ref{Fig:AND}(a), performs the same operation by flipping the target value only when both control qubits are in the $|1\rangle$ state. As the starting state of the target is $|0\rangle$, the output is $|1\rangle$ when both control qubits are in the $|1\rangle$ state, which is similar to an AND gate. A quantum X gate flips the value -- amplitude -- of its input, converting a $|0\rangle$ state to a $|1\rangle$ state, and vice versa. This operation essentially mimics a classical NOT gate, as illustrated in Figure~\ref{Fig:AND}(b). To construct a NAND gate, it is sufficient to initialize the target qubit to the $|1\rangle$ state, as illustrated in Figure~\ref{Fig:AND}(c). As a result, we obtain a NAND gate by employing an AND plus NOT gate set, which is a complete classical gate set, enabling us to represent basically all  classical logic using quantum logic. 

Quantum classical circuits are typically constructed using various $n$-qubit controlled Toffoli gates (MCT). While there also exist alternative methods for creating these circuits, they tend to be less generic or less straightforward than the Toffoli gate-based approach, and as such, they often constitute specially-designed quantum classical circuits. There are three major use-cases for quantum classical circuits:

\squishlist
    \item One reason to prefer quantum computers that support classical logic (e.g., adders, multipliers, etc) is to ensure backward compatibility with existing classical computational systems and algorithms.
    \item Another motivation is to study how classical circuits behave when interacting with quantum states as their inputs. When optimizing quantum classical circuits, via uncomputation for ancillary~\footnote{An ancillary qubit is an extra qubit in quantum circuits that has a constant input and is used to assist operations} reuse~\cite{ding2020square,parent2015reversible} (the majority of existing quantum classical circuit optimizations), this case is inapplicable due to no-cloning theorem, quantum measurement teleportation and etc.
    \item Quantum classical circuits can be utilized to perform classical operations as 'part' of general quantum circuits. Shor's algorithm~\cite{shor1999polynomial} and Grover's algorithm~\cite{grover1996fast} are two well-known examples that accommodate so called classical 'sub-circuits' -this case is not considered classical circuit optimization.
\squishend

It is important to note that quantum classical circuits have been previously researched as part of a large body of works. In fact, prior works such as ~\cite{Paler2016Resizing,ding2020square,parent2015reversible} represent only a small subset of many previously-published studies that have  utilized quantum classical circuits in various domains and settings.  

%%%%%

%% file: sections/3-related.tex
%%%%%

\section{Related Work}
\label{sec:Related}
In this section, we go over recent advancements aimed at improving the output reliability of quantum circuits. We begin by examining circuit-level optimization techniques and then proceed to discuss error minimization oriented approaches. 

\subsection{Classical Circuit Optimizations}
\label{subsec:ClassicalCircuitOptimization}
The 'size' (formally 'width') of a quantum circuit is the number of qubits it has, and the 'depth' of the circuit is the number of quantum circuit DAG levels (serial gates). Reducing the size and/or depth of a quantum circuit is important because of at least two reasons: 
\squishlist
\item {\em Improving Reliability:} As stated in Section~\ref{subsec:NISQ}, the NISQ system is primarily intended to provide reliable outputs for small-to-medium sized quantum circuits.
\item {\em Coping with Limited Resources:} Qubits are limited and valuable resources, and infinitely large quantum systems are not currently attainable. Furthermore, we currently do {\em not} have a 'quantum memory' to compensate and create the illusion of an infinitely large system, as is the case with classical computers.
\squishend
While the first reason is the primary and predominate concern -- and the main focus of our paper -- there exist approaches that target the second reason as well~\cite{sadeghi2022quantum,mmmrOctober,hua2022exploiting}. 

\begin{figure}[ht!]
\center
\includegraphics[width=.6\columnwidth]{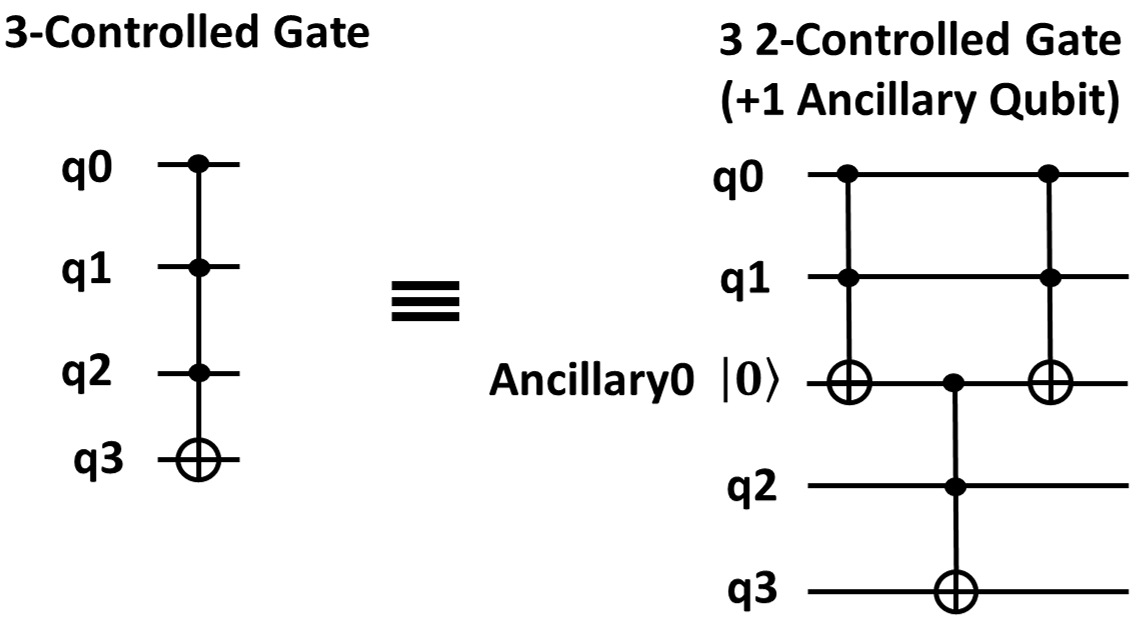}
\caption{Implementing a 3-controlled MCT gate via Toffoli gates.}
\label{Fig:anc}
\vspace{-0.15in} 
\end{figure}

Quantum classical circuits can be constructed using numerous MCT gates, but the problem is that the classical gate sets of most NISQ systems contain only the X, CNOT and Toffoli gates. To construct these various controls from the existing physical gates, we must incorporate 'ancilla-qubits', as illustrated in Fig.~\ref{Fig:anc} (other reasons for ancilla-qubits in a circuit include achieving circuit reversibility and implementing parity qubits for error correction). Reusing these ancilla-qubits improves the reliability of quantum circuits by optimizing their size and depth. Paler et al.~\cite{Paler2016Resizing} employ measurement gates to reduce the size of a quantum classical circuit via 'ancilla-qubit reuse'. In comparison, Parent et al.~\cite{parent2015reversible} utilize 'uncomputation' to reuse ancilla-qubits and reduce circuit size, thereby preserving the 'reversibility' of the circuit. Their work introduces two policies -- eagerly uncompute after use or lazily uncompute at the completion, with the former leading to smaller size with greater depth and the latter to larger size with reduced  depth. Finally, SQUARE~\cite{ding2020square} attempts to optimize circuit depth and size by focusing on the physical system architecture and selecting the most suitable uncomputation policies to produce a quantum classical circuit that is size- and depth-optimized. 

In addition to quantum classical circuit optimizations, three recent concurrent  works~\cite{sadeghi2022quantum,mmmrOctober,hua2022exploiting} attempt to reduce the size (width) of quantum circuits by employing 'middle measurement' and 'middle reset' gates (they can be applied to quantum classical circuit as well). When applicable, such circuit size reduction strategies can generate promising improvements as far as output reliability is concerned. These gains in reliability are consistent with our intuition that a smaller size and/or depth in a quantum circuit results in greater reliability in NISQ systems. Note however that, while such works try to reduce the size of a given quantum circuit, our proposed approach in this paper aims at the depth of the circuit; so, these two lines of approaches are complementary.

%%%%%

\subsection{Error Optimizations}
\label{subsec:ErrorOptimizations}
There exist various techniques for reducing each of the three categories of errors described in Section~\ref{subsec:QuantumErrors}, in order to increase output reliability. The two most important techniques for reducing coherence  errors are 'gate scheduling' and 'dynamic decoupling' (DD). Scheduling gates can be used to optimize the tradeoff between coherence errors and crosstalk errors~\cite{murali2020software}, and in particular gates can be positioned to decrease coherence errors~\cite{smith2022timestitch}. On the other hand, DD is the primary approach employed for combating coherence errors. Pulse control that reduces the qubit's frequency to increase T1/T2 and gate string repetitions~\cite{10.1145/3466752.3480059,smith2022timestitch} (such as XY,  X, ...) with a delay in between to frequently reduce the t/T variable in the exponent are the two main incarnations of DD. 

Crosstalk and gate errors can be addressed to some extent via 'logical-to-physical qubit mapping' as well as 'gate cancellation'~\cite{li2022paulihedral,Liu2022not}. SWAP gates increase, in the absence of a link, the circuit depth;  therefore, mapping techniques attempt to reduce the number of SWAP gates to be added to the circuit~\cite{Liu2022not,zhang2020depth,li2019tackling,tannu2019ensemble,liu2021qucloud,dou2020new}, which in turn reduces gate errors and potential coherence errors. In comparison, gate cancellation reduces the number of gates and can offer similar advantages~\cite{li2022paulihedral,Liu2022not}. By carefully positioning qubits and routing the SWAP operations in directions that avoid crosstalks, mapping can prevent/reduce crosstalk errors~\cite{Khadir2023TRIM,ohkuracrosstalk,liu2021qucloud}. Our work in this paper differs from these prior reliability-centric efforts in that our focus is not merely on increasing reliability, but rather on ensuring  reliability 'regardless of the circuit depth' through noise filtering, to  enable unrestricted circuit depth expansion. It's crucial to mention that our methodology complements these optimization techniques rather than overlapping with them. When implemented concurrently, our approach may result in fewer jobs or an increased block size. Also, our work reduces the algorithmic complexities of these works to 'linear time' if it is used in conjunction with them. We will elaborate on these aspect more in Section~\ref{sec:Design}.  

%%%%%

%% file: sections/4-motivation.tex
%%%%%

\section{Motivation}
\label{sec:Motivation}
This section begins with a discussion of our problem definition and the severity of the problem. It then proposes a solution based on our observations and the fundamentals of the NISQ systems. 

\begin{figure}
\center
\includegraphics[width=0.9\columnwidth]{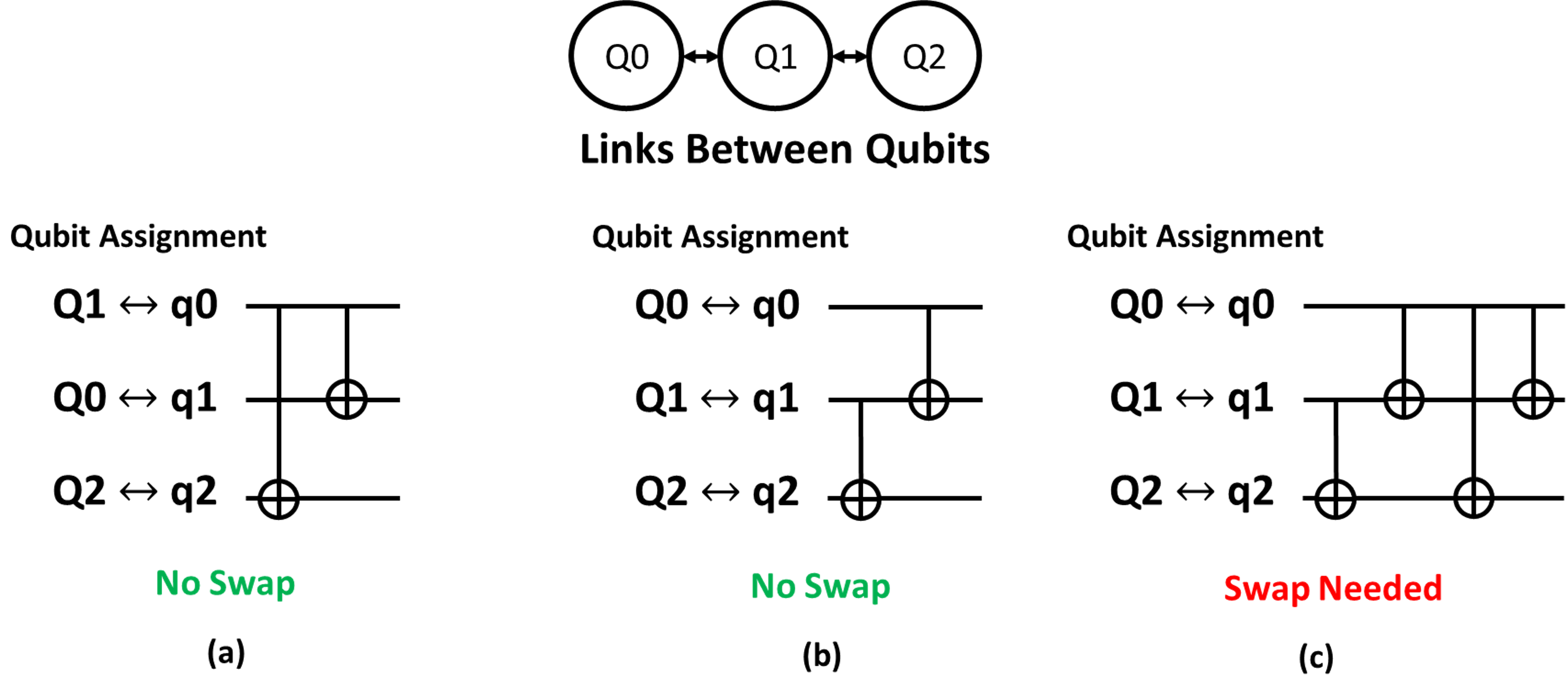}
\caption{SWAP addition following the circuit-to-architecture mapping for three different cases.}   
\label{Fig:MapMotiv}
\vspace{-6mm} 
\end{figure}

\subsection{Reliability of Large Quantum Circuits}
\label{subsec:ReliabilityofLargeQuantumCircuits} 
As vendors increase the scale of their quantum machines and systems~\cite{IBMQ}, it becomes more likely that larger quantum circuits will be executed on them. Accordingly, the reliability of the outputs generated by such circuits declines to alarming degrees. It is important to note that, for obtaining a reliable output, two criteria must be met -- (i) the probability of achieving the correct output should be the highest, and (ii) there should be a distinguishable difference between the correct output and the second (false) output. For instance, having the correct output with a 40\% probability while obtaining the second (false) output with a 39\% probability does not indicate a reliable output, as it is difficult to determine which result is accurate. Among the errors enumerated and discussed in Section~\ref{subsec:QuantumErrors}, gate errors are the least problematic ones. In fact, the main idea behind dynamic decoupling (DD) optimizations \cite{das2021adapt,bylander2011noise,khodjasteh2005fault,khodjasteh2007performance,niu2022analyzing,mena2023protectability} is to {\em convert} coherence errors to gate errors. For example, assuming a circuit of size 10 qubits with an average gate count of 40 gates per qubit line -- a total of 400 gates after transcompile -- on a system with an average gate error rate of $1\%$, we suffer an average error rate of nearly $40\%$ due to gate errors alone. Per the SABRE~\cite{SABRE} mapping and routing policy, which is a commercially-available optimized qubit mapping and routing policy, the execution of a Multiplier\_n25~\cite{li2020qasmbench} (a 25-qubit circuit) circuit on the Ibmq-montreal~\cite{IBMQ} machine resulted in a PST~\footnote{PST, or Probability of successful Trial, is the fraction of the number of correctly measured outputs within all measured outputs.} of less than $2\%$ which is practically 'noise', {\em not} an 'output'! 

Let us now illustrate the issue with a similar problem and its solution in a classical computer. It is well-known that a DRAM cell can store data for a limited time before the capacitors discharge and the data gets erased (due to leakage). To prevent this potential data loss in DRAM, we need a mechanism that modifies the DRAM cell before this threshold is reached.
Now, we can apply any read/write optimization policy to the cells, but one eventually faces an insurmountable obstacle -- program access patterns. Eventually, the solution would be determining the retention times of different DRAM rows and refreshing the rows' stored values to retain them and hence avoid leakage error. This approach is straightforward, yet effective. By employing error correction codes, we have the ability to tolerate a degree of errors, effectively extending the retention time and thereby reducing its associated costs. However, we cannot completely negate the retention time through error correction. This is primarily due to the fact that error correction mechanisms are incapable of compensating for a complete loss of data. Note that, producing reliable results from large quantum circuits presents a similar challenge here. The NISQ systems are inherently noisy, and regardless of the optimizations we employ to increase their output reliability, we will ultimately reach a 'wall'. To enable circuit size/depth to grow infinitely, we need a mechanism {\em analogous} to the DRAM retention time, like, for example, a 'qubit retention time' mechanism or, more accurately, a 'noise filtering mechanism'. Similar to the DRAM retention time, such a mechanism targeting quantum circuits should be simple, effective, and universally applicable. In this context, quantum error correction codes can help but can't eliminate the need for this mechanism, as they can't compensate for total data loss.

\subsection{Mapping and Algorithm Complexity}
\label{subsec:MappingAlgorithmComplexity}
Numerous mapping~\cite{Liu2022not,zhang2020depth,li2019tackling,tannu2019ensemble,liu2021qucloud,dou2020new,ohkuracrosstalk} and gate scheduling~\cite{murali2020software} methods have been proposed for NISQ systems over the past decade, aiming to optimize circuits for the underlying hardware. The main driving force behind these techniques is the recognition that each qubit possesses unique characteristics (such as coherence error and number of connections), and to achieve the most accurate results, these methods must utilize the most reliable qubits while minimizing CNOTs and coherence errors. Mapping techniques often involve significant graph processing, with complexities reaching up to factorial levels~\cite{tannu2019ensemble,Liu2022not}. As a result, as the circuit's width/depth increases, the algorithm complexity and execution time grow exponentially. This observation is also confirmed by~\cite{tannu2019ensemble}, which states that qubit mapping is an NP-complete problem. Consequently, obtaining the most optimal mapping is not always feasible for large circuits or systems, and as a result, most existing techniques typically settle for suboptimal mappings. 

Furthermore, conventional mapping algorithms and gate scheduling approaches demand resource-intensive 'error profiling' for physical qubits and their connections. This process requires 'system calibrations' from vendors on a {\em daily basis} to generate relevant statistics, rendering the device under test virtually inoperable during those periods~\cite{IBMQ,murali2020software}. Employing small quantum circuits not only circumvents the NP-complete algorithm input size, yielding super-linear performance gains, but also enables the production of verifiably-reliable outputs without necessitating mapping optimizations or error profiling.  

Figures~\ref{Fig:MapMotiv}(a) and (b) depict circuits that can be mapped to the underlying quantum system without necessitating any additional SWAP operations. In contrast, Figure~\ref{Fig:MapMotiv}(c) illustrates a circuit that indeed requires a SWAP operation, even though it is formed by simply {\em merging} the circuits in parts (a) and (b) together. That is, while the small circuits in parts (a) and (b) individually do {\em not} require any SWAP operation, a straightforward combination of them in part (c), which is relatively larger than (a) and (b), {\em does}. Hence, this example illustrates that by running smaller circuits, the output reliability can be   enhanced, aligning well with the objective of this paper. This observation about the NISQ systems can be generalized: 
\begin{quote} 
{\em The smaller the quantum circuit, the less likely it is that additional SWAPS would be introduced; consequently, we would  experience better execution time, lower cost, and improved output reliability with smaller circuits.} 
\end{quote} 
The  technical details of our proposed design are given later in  Section~\ref{sec:Design}.

\begin{figure*}[t!]
\center
\includegraphics[width=0.7\textwidth]{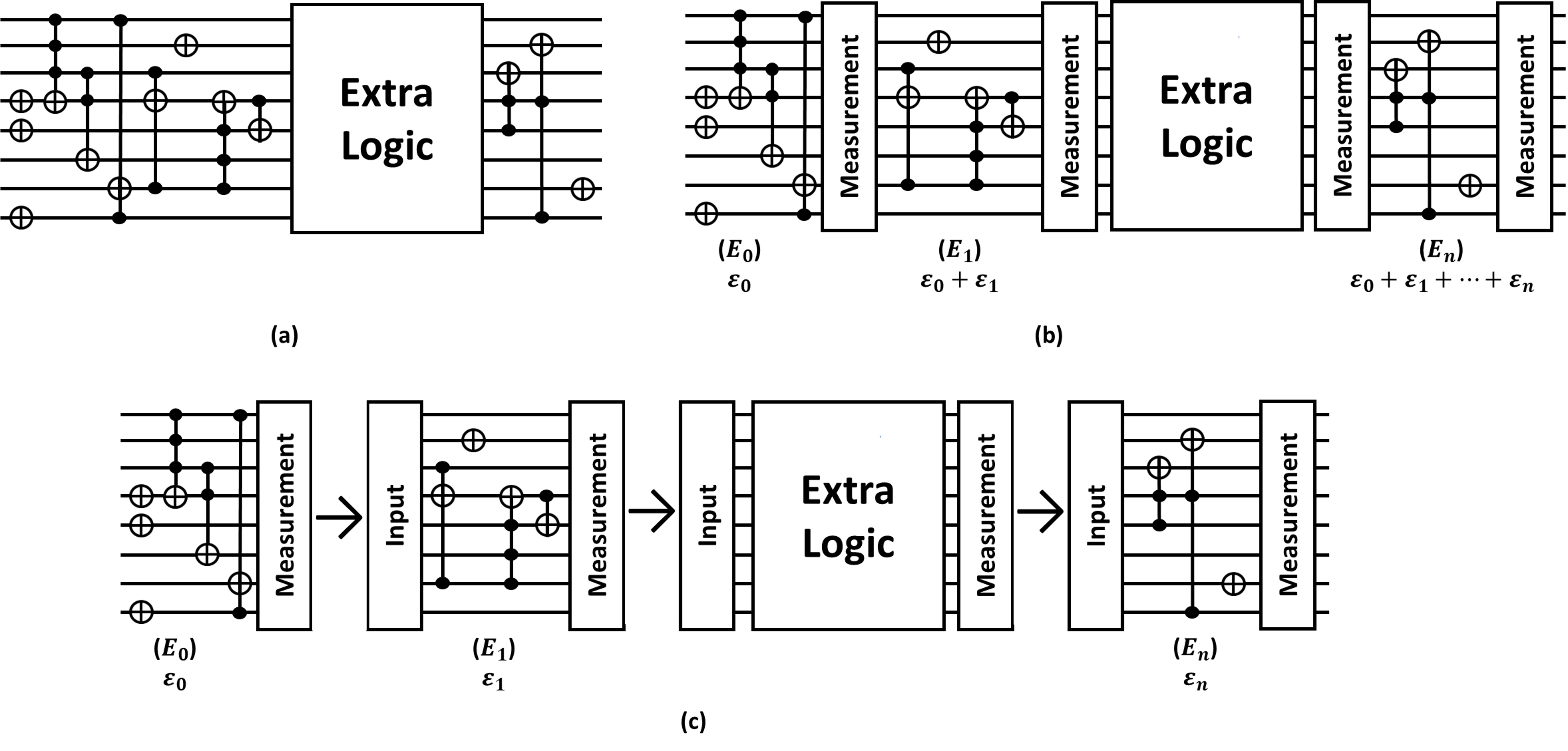}
\vspace{-0.1in} 
\caption{(a) An infinitely-deep quantum classical circuit; (b) A microscopic look at the stages of the circuit. Adding only 'measurement gate' with {\em no} noise filtering mechanism; and (c) The modified circuit with noise filtered.}
\vspace{-6mm}
\label{Fig:Slice}
\end{figure*}

\subsection{Motivation Summary}
\label{subsec:MotivSum}
To recap, the goal of this study is to develop a method to satisfy the following:  
\squishlist
\item First and foremost, we would like to develop a simple yet effective universal method, specifically for quantum classical circuits where a supportive technology, namely, classical memory,  already exists. This method would also provide insights for quantum circuits in general when quantum memory/ quantum noise filters become available. Our proposed approach, {\bf Depth Control} enables us to infinitely increase the size/depth of circuits while maintaining the reliability of the results on a NISQ system.
\item Secondly, we intend to make it possible for more optimized mapping algorithms to be applied to the target circuits by obtaining 'super-linear' speedups by reducing the input sizes of the NP-complete mapping algorithms. 
\item Thirdly, we want our method to be as 'fail-safe' as possible, so that we have confidence in the reliability of any large circuit even {\em without}  mapping, or gate scheduling, or any other optimization, thereby eliminating the need for (costly) system calibrations from vendors. 
\item Lastly, we intend to identify potential opportunities for existing mapping algorithms to reduce the number of SWAPs in the final executable circuit, thereby reducing the circuit execution latency and lowering the incurred costs. 
\squishend

%%%%%

%% file: sections/5-design.tex
%%%%%

\section{Design and Implementation}
\label{sec:Design}
In this section, we first outline the observations that form the basis of our proposed design. Next, we provide a detailed description of the design and justify it based on the stated observations. Following that, we address a potential issue that may arise from the proposed strategy and offer a solution for it. Lastly, we assess our two proposed strategies and discuss their applicability and effectiveness.

\subsection{Heuristic and Overall Goal}
\label{subsec:Heuristic}
First, let us identify the root cause of the problem. Figure~\ref{Fig:Slice}(a) illustrates a typical quantum circuit of medium size with an unknown depth and number of gates. The errors at the end of each stage are represented by $E_0=\epsilon_0$ to $E_n=\epsilon_0+\epsilon_1+...+\epsilon_n$. It is clear that $E_n=\epsilon_0+\epsilon_1+...+\epsilon_n$ is greater than $E_n-1=\epsilon_0+\epsilon_1+...+\epsilon_n-1$, and $E_1$ is greater than $E_0=\epsilon_0$, as errors accumulate and increase with the addition of each gate or stage. Moreover, noise does not decrease with the addition of a gate, as negative noise does not exist in nature. As the number of gates increases, we observe that $E_n\geq 50\%$ reaches a point where noise simply becomes {\em dominant}, leading to practically 'unreliable' results.

Figure~\ref{Fig:Slice}(b) provides a close-up view of the circuit shown in Figure~\ref{Fig:Slice}(a). If we can eliminate errors and purify the states at the {\em end} of 'each block' (as indicated by the arrows in Figure~\ref{Fig:Slice}), by correcting their states and storing them in a memory, we can ensure a certain level of 'reliability' at the {\em end} of the entire circuit. However, this approach would necessitate the use of a quantum memory, which is currently not available. On the other hand, we do have classical memory that can work in tandem with quantum systems, allowing us to potentially handle the quantum classical circuits, if not the generic quantum circuits. Consequently, our focus will be on the quantum classical circuits, as we aim to address the problem within this context.

The goal here is to {\em correct} the states of the qubits at the end of each block shown in Figure~\ref{Fig:Slice}(b). Since it is a quantum classical circuit, the state of each qubit in a perfect, noiseless system should be either $|0\rangle$ or $|1\rangle$. However, due to noise, the state of each qubit at the end of a block is either $(1-\epsilon)|0\rangle+\epsilon|1\rangle$ or $(1-\epsilon)|1\rangle+\epsilon|0\rangle$, deviating from the ideal $|0\rangle$ or $|1\rangle$ states. By performing a measurement, these quantum states will be collapsed and recorded as either $0$ or $1$, based on their expected values. Consequently, the states are restored, allowing us to proceed to the next stage. While, at first look, it seems, by merely applying a measurement gate (Fig.~\ref{Fig:Slice}(b)), the classical state would recover automatically, in reality doing so would just {\em add}  a 'gate measurement noise' into the existing noise. For instance, consider a situation where, after applying a measurement gate to an expected $0$ value, the noisy state converges to $0$ (with a probability of $1-\epsilon$) and to $1$ (with a probability of $(\epsilon)$). After the measurement, we would get $(\epsilon+\epsilon_m)\times{n}$ measured values of $1$ -- $n$ being the number of iterations/shots~\footnote{A 'shot' refers to a single execution of a quantum circuit on a quantum processor, allowing for the collection of measurement results and the exploration of probabilistic outcomes.} of circuit execution -- and $(1-(\epsilon+\epsilon_m))\times{n}$ measured values of $0$ solidify the noise error and introduce a measurement noise into the original circuit. Therefore, simply applying a measurement gate alone (similar to the DRAM retention time) would {\em not} rectify the state but would instead simply  increase the error from $\epsilon$ to $\epsilon+\epsilon_m$. 

Since using the measurement gate alone is not sufficient to achieve our desired goal, we need to incorporate a {\em mechanism} to filter out noise and purify the results at the end of each block. As illustrated in Fig.~\ref{Fig:Slice}(c), if $\epsilon_i$ at the end of each block is small enough, we can reasonably assume that the correct output would be stored with the highest counts. To ensure that $E_i$ is small enough, it is necessary to ascertain that the blocks are small enough so that the probability of obtaining an incorrect result is minimal. By selecting a reliable small block, we can confirm that the correct output is measured with the highest probability, which we will then use as 'input' for the following block. In other words, if the output with the highest count at the end of the $(i_{th})$ block is the $(001110101...)$ stream, then the input for the $(i_{th}+1)$ block is the $(--xxx-x-x...)$ gates. As a result, the errors at the end of each block are eliminated, and the final (result) error is limited by the error of the last block, as opposed to the accumulation of the errors from all the blocks. Therefore, we reach the following important {\bf conclusion}:  
\begin{quote}
{\em If all blocks are 'sufficiently small', the quantum classical circuit can be infinitely large (in depth, not size, as size/width is constrained by the number of qubits in the system) with a bounded error equal to the error of the last block $(\epsilon_n)$ on any NISQ system.}
\end{quote} 

Our main objective is to determine the block size that enables us achieve our goal of {\em limiting} error propagation for an infinitely-large circuit. Furthermore, we aim to facilitate additional mapping optimizations by creating blocks before transpiling, allowing the transpiler to use any level of optimizations for the smaller sub-circuits. It is important to note that this block creation could be performed after transpiling, but this comes at the cost of losing mapping optimizations, a mentioned in Section~\ref{subsec:MappingAlgorithmComplexity}. In the evaluation section, we explain why it is not recommended to forgo existing optimizations and rely solely on DC. Our proposed DC can be implemented in two different ways -- static depth control and dynamic depth control -- which are discussed next. 

\subsection{Static Depth Control (SDC)}
\label{subsec:SS} 
In this section, we introduce our first approach, Static-Depth-Control (SDC), keeping two primary objectives in mind: i)  {\em guaranteed reliability of each block}, and ii) {\em simplicity and applicability of the algorithm}.

In the {\em Static-Depth-Control} (SDC) approach, our goal is to create 'uniformly-sized' blocks with a high probability of obtaining the correct outputs. Ensuring the accuracy of the outputs is of utmost importance, as the objective is to filter out the 'noise' to get the 'correct output'. In other words, if an error occurs in the initial or intermediate layers, it can propagate to the end of the circuit, leading to an incorrect result being reported with high confidence and count. Now, we need to identify the 'slicing criteria' that can ensure minimum error on the sliced circuits. We consider two primary slicing strategies for our circuits: i) {\em slicing based on the number of gates from the circuit:} this approach slices the circuit in a way that ensures a similar number of gates in different blocks; and ii) {\em slicing  based on the number of layers in the directed acyclic graph (DAG):} this approach slices the circuit such that each block has a similar number of DAG layers.

Although slicing based on the number of layers in the DAG appears promising, this approach is clearly circuit-dependent and may result in one block containing multiple high-error gates while others do not. As a result, we opt for slicing based on the gate count, as it offers greater reliability and leads to a more even distribution of blocks. Further, as the system grows larger, there are fewer and fewer links between qubits~\cite{IBMQ}. For example, IBM's largest advertised system has 433 qubits; so, a system with 400 qubits could hypothetically have a DAG layer with 100 CNOT operations (each CNOT is on 2 qubits $2\times100=200$ can fit easily in 300 qubits to execute in parallel). Even if we slice on the premise of a single DAG layer, the number of potential SWAPs -- due to low link count and number of CNOT gates -- and the possibility of crosstalks would render the output unreliable. Therefore, we believe that the safest and more balanced method would be slicing based on gate counts. 

\RestyleAlgo{ruled}
\begin{algorithm}[t]
\scriptsize
\caption{DDC algorithm.}\label{alg:DS} 
\textbf{Input:}
\BlankLine
$Circuit$
\BlankLine
$n\:\longrightarrow$ number of gates in circuit
\BlankLine
$R_{Threshold}\:\longrightarrow\:$ the minimum acceptable fidelity of $circuit[i]$ block
\BlankLine
\textbf{Register:}
\BlankLine
$Fidelity = 0$
\BlankLine
$Result = null$
\BlankLine
\textbf{Output:}
\BlankLine
$Final\_Resaults$
\BlankLine
\textbf{Algorithm:}
\BlankLine
$Circuit[0]=Circuit$
% \BlankLine
% \eIf{}{}{}
\BlankLine
\For{i \textbf{in} \em{range(n)}}{
$Circuit[i] \longleftarrow Circuit.break(\frac{n}{i+1})$ \tcp*{grab the first $\frac{n}{i+1}$ gates in the circuit}
\BlankLine
$Fidelity = Circuit[i].fidelity()$ \tcp*{calculate the fidelity of circuit}
\BlankLine
\If{($Fidelity\:\geq\:R_{Threshold}$)}{
$Result = Circuit[i].measure()$
\BlankLine
\For{ i,yes \textbf{in} enumerate(reversed(Max(Result))}{
    \If {yes == '1'}{
        Circuit.x(i)
        }
        } \tcp*{for 1s in key with max count we add an X gate as initialization for the circuit remnants}
\BlankLine
$Circuit = Circuit - circuit[i]$ \tcp*{attach the remnant of the circuit to the initialization}
\If{$Circuit==null$}{$Final\_Resaults = Result$}
} } 
\vspace{-0.1in} 
\end{algorithm}

Now that our slicing criteria has been decided, we can proceed to explain the details of our Static-Depth-Control (SDC) strategy. We define a {\em constant} 'block size', based primarily on the desired system attributes. This size must be small enough to ensure that we can confidently obtain the correct output, regardless of the executed circuit. Theoretically, this dimension could be as small as a single gate, the smallest meaningful circuit conceivable, if the system is highly noisy and generally unreliable. Clearly, a system that fails to produce the correct output even for a single-gate circuit is neither dependable nor practical. The SDC strategy can be described as follows:

\begin{figure}[t!]
\center
% \vspace{-4mm}
\includegraphics[width=\columnwidth]{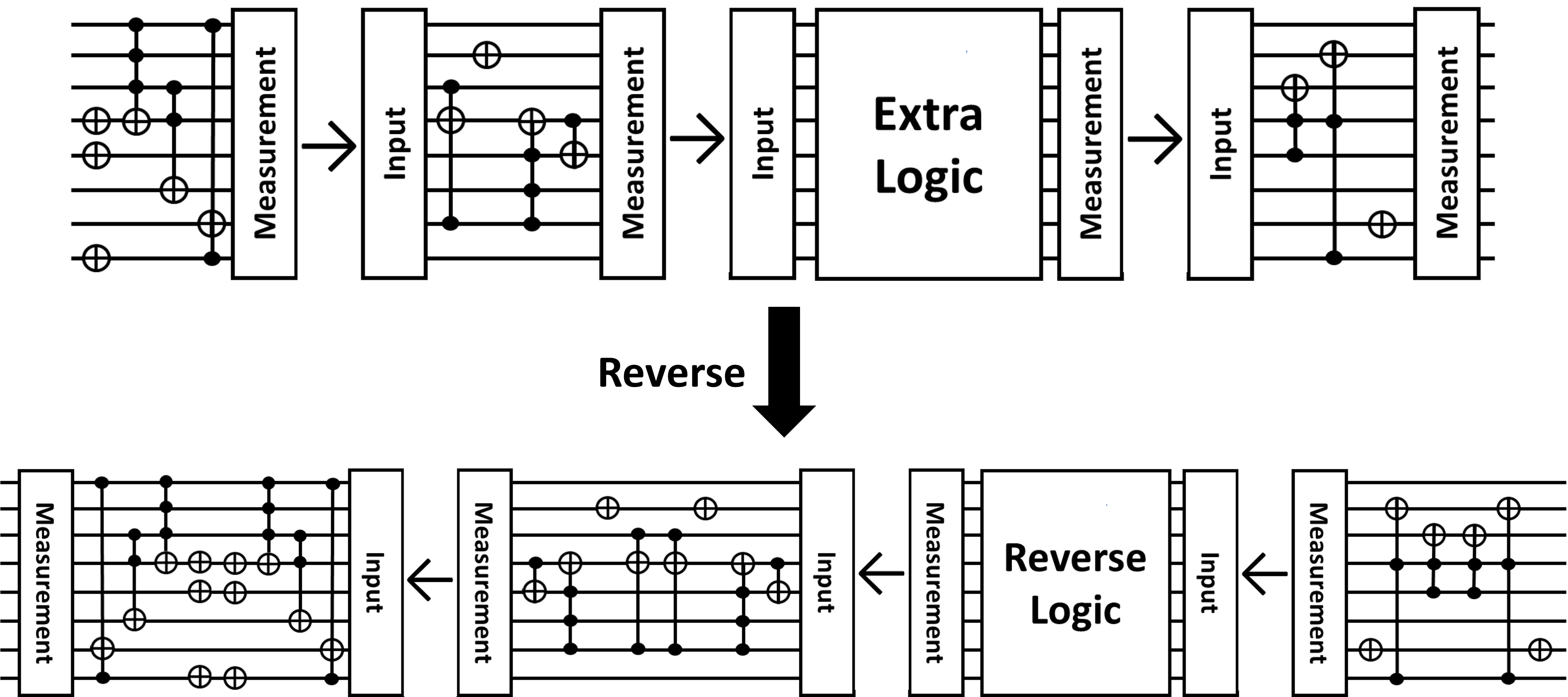}
\caption{Illustration of the 'reversing-procedure' on the DC generated circuit.}
\vspace{-6mm}
\label{Fig:Reverse}
\end{figure}

\squishlist
\item First, we start by counting the total number of gates. In our approach, we do not consider a unitary gate (in a quantum classical circuit, an X gate) as part of the total gate count. 

\item Each of the basic controlled gates that are physically supported by the underlying hardware counts as 'one gate' in our total gate count calculation.

\item Each gate with more than two control signals is counted as $1 + 2 \times (n - 2)$, where $n$ represents the number of control signals in the gate. It is important to note that implementing these gates using CNOTs or CCNOTs requires one ancilla qubit, with a computation and an uncomputation gate for each additional control signal. If the total number of gates generated from this operation exceeds our gate count 'threshold' ($d$) -- recall that we are slicing based on gate count --, we compile it into basic gates and apply the slicing technique to it as well. 

\item Upon the completion of each block's execution, the measured output with the highest number of counts will be used to {\em initialize} the subsequent block. This initialization can be accomplished by applying X gates to each qubit, if the input is expected to be 1.

\item Depending on $d$ and the gate count of the circuit, the final block is either as large as $d$ or smaller. 

\squishend

To recap, our SDC method is characterized by its low algorithmic complexity. However, it may result in a substantial number of 'sub-circuits' (or 'jobs'), which will be further discussed in Section~\ref{subsec:DesignDiscussions}.

\subsubsection{SDC effect on mapping optimizations}
Recall that one of our objectives was to facilitate optimizations on the complexity of advanced mapping/routing algorithms for all quantum-classical circuits. In this section, we discuss the impact of DC on the complexity of these algorithms. The worst-case timing for a mapping and routing algorithm on a circuit with $n$ active qubits~\footnote{Active qubits are qubits with at least one assigned gate on them (usually control signal gates)} and $m$ gates that consider all the different combinations is equal to $O(m!n!)$~\cite{tannu2019ensemble}. Using our algorithm, instead of mapping and routing a single circuit containing $n$ qubits with $m$ gates, we need to perform the mapping/routing $\ceil{\frac{n}{d}}$ times (number of blocks). In each mapping, DC has at most $d$ gates (block size of the SDC), and the number of active qubits in each circuit is equal to $3d$ since each gate in our technique can have at most two control signals (CCNOT) and the highest number of active qubits they can use is equal to $3d$. Therefore, the worst-case timing complexity for each mapping would be equal to $O((3d)!d!)$, which results in an overall complexity of our algorithm equal to $O(\ceil{\frac{n}{d}}\times{((3d)!d!)}$. Since $d$ is constant in our algorithm, the worst-case timing complexity of our algorithm will be equal to $O\ceil{\frac{n}{d}}\times((3d)!d!)=O(kn)=O(n)$. Thus, while the original mapping algorithm has a worst-case execution time of $O(m!n!)$, DC can reduce it to $O(n)$.

%One of our motivations was to facilitate optimizations of sophisticated mapping/routing algorithms for all quantum classical circuits. This section discusses the complexity of these optimizations following the deployment of DC. Let's presume the worst-case exponential growth complexity of $\mathcal{O}(N^N)$ and analyze the algorithm's complexity after DC application for the big $\mathcal{O}$ notation analysis of our SS approach effect on the circuits. if we suppose a circuit with $m$ qubits and $n$ gates, then $N = m\times n$. If the slice size to guarantee reliable output is $d$ number of gates, then the largest value of $m$ (width/number of qubits) for each slice will be ($(3d+1)d$ qubits on that slice (the widest gate that contribute as $1$ count to d is CCNOT with 3 qubits width --> $((3d+1)d)$ -- we can agree constant $d$ limits both the size and gate counts to a constant value --) and we will have $\frac{n}{d}$ circuits to optimize. Therefore each bock has $Max(m)=(3d+1)d)$ and $Max(n)=d$ and we will have $Max(N)=((3d+1)d)\times d)$. Each block's optimization has a maximum complexity of $\mathcal{O}(k=((3d+1)d)\times d)^{((3d+1)d)\times d))})$, and since size $d$ is a constant in our SS approach, $k$ will also be constant, so its complexity is $\mathcal{O}(k)=\mathcal{O}(1)$. We have $\frac{n}{d}$ circuits, so we will need to optimize $\mathcal{O}((k/d)n)$ circuits; therefore, the new complexity of these algorithms will be $\mathcal{O}(n)$, regardless of their original complexity.}
%%%%%

\subsection{Dynamic Depth Control (DDC)}
\label{subsec:DS} 
In an attempt to decrease the number of jobs generated by the sliced circuit, we can enlarge the size of each block, accepting a higher level of output 'reliability risk'. Given that the job queue operates on a Longest Job First (LJF) policy~\cite{ravi2021adaptive}, employing a more intricate algorithm may help reduce waiting time while sacrificing some reliability. It is important to note that the circuits targeted in this study are initially large enough to yield unreliable outputs, making this tradeoff reasonable. To that end, we propose Dynamic-Depth-Control (DDC), which is essentially a 'divide-and-conquer' approach, designed to establish the appropriate block sizes.

We want to emphasize that, unlike the SDC method, the DDC approach allows for 'varying' block sizes, based on the gate operations within each block. Although DDC generally takes longer to generate its circuit, the resulting circuit is expected to produce its output faster. Both DAG layers and gate counts can be utilized in the DDC algorithm, with gate counts potentially providing greater accuracy. For the sake of consistency with the SDC method, we use 'gate counts' in our algorithm; however, one may  opt, if desired, for DAG layers (in the DDC approach).

Our DDC divide-and-conquer strategy is presented in Algorithm~\ref{alg:DS}. The process of transmitting data from one block to another is the same as in the SDC scheme. The main difference lies in the 'selection of blocks', which can have varying sizes as long as they meet the user-defined reliability threshold. In each iteration, we divide the circuit in {\em half} and evaluate the block's fidelity. If the fidelity satisfies the threshold, we proceed to use the same counting standard specified in SDC~\ref{subsec:SS}, initializing the remainder of the circuit with the output of the previous iteration. If the fidelity is below the threshold, we continue to halve the block until the 'desired fidelity level' is achieved. Calculating the fidelity of a circuit on the target hardware is a more complex operation compared to using constant-sized blocks in SDC, but it allows for dynamic block size expansion when possible, reducing the number of jobs.

It is crucial to maintain a 'sufficiently high' threshold, as simulations often {\em overestimate} the actual output reliability due to the inability to accurately account for certain errors like crosstalk. Furthermore, the more accurate the simulation needs to be, the longer it takes to generate the final sliced circuit, due to the compilation time. A higher threshold leads to a lower error rate and a larger number of jobs. Conversely, a lower threshold would result in fewer jobs at the expense of accepting a higher error risk.

\subsection{Addressing the Reversibility of Quantum Circuits after Applying SDC/DDC}
\label{subsec:reversibility} 
Introducing a measurement gate into a circuit raises valid concerns about the 'reversibility' of the circuit~\cite{Paler2016Resizing, ding2020square}. Fig.~\ref{Fig:Reverse} illustrates how our approach can {\em preserve} reversibility between blocks. While each block remains a fully-reversible quantum classical circuit to which the same uncomputation technique can be applied, the primary concern lies in the reversibility of the {\em entire circuit}. To achieve uncomputation in a quantum circuit, we need to add inverse gates in reverse order. Alongside the reverse addition of gates, we must also execute the DC algorithm in {\em reverse}. This would result in both reverse gates and a measurement gate to be added into the opposite side, serving as input for the uncomputation logic. Fig.~\ref{Fig:Reverse} helps to visualize this approach, showing that a certain level of reversibility can be maintained, with the main change being the propagation of measurements through the blocks -- similar to the DC propagation, but in the reverse direction. Further optimizations for uncomputation remain a subject for future work, but as far as this paper is concerned, the reversibility of the circuit is maintained after applying the DC method.

%%%%%
\subsection{Slicing Policy for Superposition States of Inputs}
%%%%%
Previous works on quantum-classical circuits typically assume  classical values as inputs. However, when these inputs are in a 'superposition state', a quantum-classical circuit begins to mirror a general quantum circuit, thereby invoking all limitations of quantum theory such as the no-cloning theorem and the quantum measurement teleportation. More specifically, many prior quantum-classical optimizations, such as those detailed in~\cite{ding2020square,parent2015reversible}, are  unsuitable for superposition states of inputs due to the no-cloning theorem, while the work of Paler et al.~\cite{Paler2016Resizing} is unsuitable (again, for superposition states of inputs) due to quantum measurement teleportation. In this study, {\em we propose a technique capable of functioning with superposition states}, thus aiming to create a more generalized strategy. Furthermore, we intend to foster the development of 'quantum filters' for quantum memories to reduce the overall system error, especially when the existing error correction techniques fall short. In numerous scenarios, neither present nor future quantum error correction strategies will be able to completely eliminate errors, highlighting the advantages of our proposed technique even in the long run.

In DC, we also strive to produce reliable outputs through circuit slicing when dealing with superposition state inputs. Commonly during the initialization, the input is positioned into a state encompassing multiple classical inputs (ranging from $2$ to $2^n$), each with varying probabilities. Let us assume that our initialized state is $|\psi\rangle=\sum_{i}\alpha_i\times|i\rangle$, which can yield result  $i$ with a probability of $|\alpha_i|^2$ following measurement. Thus, if we perform an experiment with 10000 shots ($ns=10000$), we understand that $ns\times|\alpha_i|^2$ experiments would output $i$. While  theoretically a system can contain $2^n$ states, the finite number of shots we can execute limits the attainable states. To account for a potential margin of error in outcome measurements, a typical maximum  number of input states in practice would typically be less than or equal to $n$ states and significantly less than 10000 -maximum number of shots for IBMQ. Bearing these considerations in mind, we propose a feasible method for applying DC to the input superposition states under two constraints:
\squishlist
\item Upon the specified initialization circuit, the programmer must provide the anticipated number of states, but not the actual outcome or expected distribution of outcome.
\item Due to the indistinguishable state post measurement problem (e.g.,  $|+\rangle$ would produce the same measured outcome as $|-\rangle$), the reversibility of the circuit cannot be maintained using the currently-available technologies while applying DC on superposition inputs.
\squishend 

\begin{figure*}
\center
\vspace{-3mm} 

\includegraphics[width=0.7\textwidth]{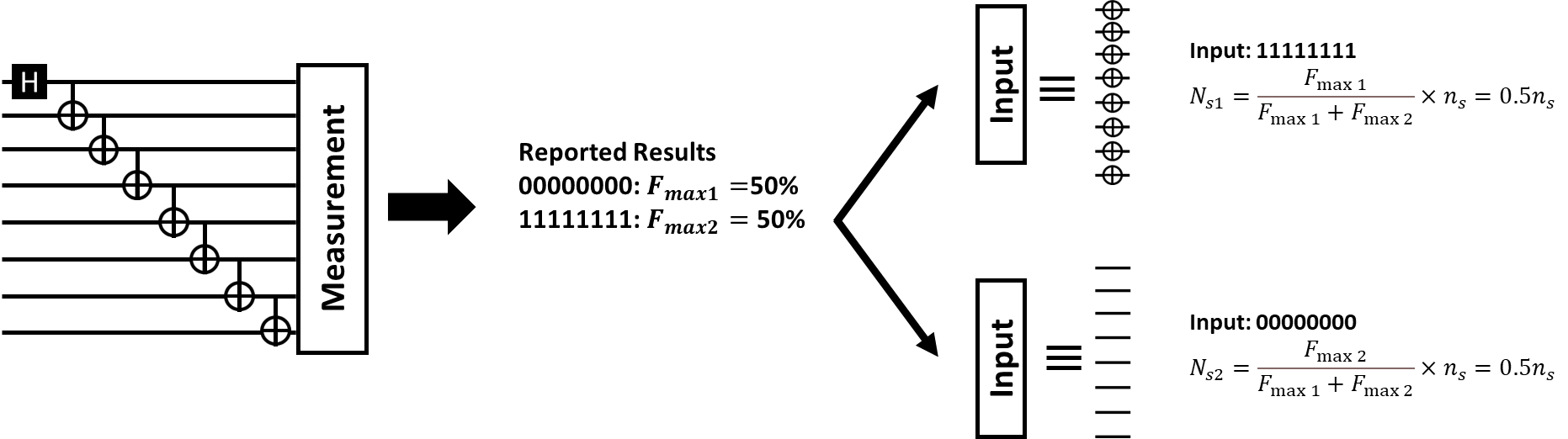}
\caption{DC application on superposed inputs}   
\label{Fig:Superposition}
\vspace{-6mm} 
\end{figure*}

Once the number of anticipated outputs ($k$) is established and the initialization is complete, the top $k$ most frequently reported results after the initialization are selected as the 'expected outputs', with the remaining outputs filtered out as 'noise'. Note that the identification of the initialization phase is straightforward when the $initialize()$ function in Qiskit~\cite{Qiskit} is employed, but a manual initialization requires the programmer's input to identify the initialization circuit. We further presume that the most frequently reported outputs align with the expected outputs, if the initialization circuits are small enough and unlikely to incur significant errors. Consequently, for each expected output, we input it into the classical circuit and execute it on the hardware, as depicted in Fig.~\ref{Fig:Superposition}(b). The number of shots for each circuit is computed using the formula $Shots[i]=\frac{N_i}{\sum_{1}^{i}N_i} \times n_s$, where $n_s$ represents the total number of shots and $N_i$ is the number of observed results for output $i$, post-initialization. Therefore, we distribute the number of shots based on the likelihood of observing $i$ in the initialization phase. The subsequent step involves producing the final results based on the output from each and every circuit. This involves collating the number of observed results from each circuit into a singular dictionary for user reporting. As Section~\ref{subsec:Heuristic} discusses, slicing the circuit decreases errors for each initialized circuit, thereby yielding more precise outputs than their non-sliced counterparts even with inputs in superposition states.

To prove the correction of DC under superposition inputs, we assume a 'noiseless ideal system' and measure the output at the end of the circuit (original case). Then, at the conclusion of the initialization, we apply a measurement, followed by the creation of the distributed classic inputs to the circuit and concatenate the end results after the final measurement (DC). We then show that both outcomes are {\em identical}. 
\squishlist
\item Original case:
$|\psi\rangle=\sum_{i}\alpha_i\times|i\rangle$$\Rightarrow|\langle\phi|\psi\rangle|^2$$=\{E_1:n_1,E_2:n_2,...,E_i=n_i\}$ where $n_i=|\alpha_i|^2n_s$ and $E_i=T(i)=\lambda_i|i\rangle$, 
where $\psi$ is the input state, $\phi$ is the quantum classical circuit matrix, and $\lambda_i$ are the eigenvalues for $\phi$. 
\item DC case:
$|\psi\rangle=\sum_{i}\alpha_i\times|i\rangle$ by applying a measurement gate right here we will have $\{\langle1|:n_1,\langle2|:n_2,...,\langle i|:n_i\}$ where $n_i=|\alpha_i|^2n_s$. Feeding each $|i\rangle$ vector as an input into the classical circuit for $n_i$ shots where $\sum_i n_i=n_s$ and concatenating them into one dictionary, we have the post measurement outcome as $\{\lambda_1|1\rangle: n_1,\lambda_2|2\rangle:n_2,...\lambda_i|i\rangle:n_i\}=\{E_1:n_1,E_2:n_2,...,E_i=n_i\}$ which shows that, under both the cases,  the outcome is the {\em same} on an 'ideal system'. 
%Although, on the noisy NISQ due to noise filtering mechanism of DC the reliability of outcome is guaranteed unlike the original case.
\squishend

\subsection{Discussion}
\label{subsec:DesignDiscussions}
In this section, we compare our previously-proposed DC strategies -- DDC and SDC -- which involve manipulating the (classical) memory to limit the errors of any large (classical) quantum circuit to a limited size circuit block ($\epsilon_n$)~\ref{subsec:Heuristic}. Recall that, DC is essentially a noise filtration mechanism and a value propagation strategy, which can be implemented {\em both} statically (SDC) and dynamically (DDC). 

In the DDC method, circuits are sliced into equal-sized blocks with a chosen number of gates, as small as desired, to achieve a conservative and reliable result. In a 'worst-case'  hypothetical system, the block size could be as low as 'one gate'. The issue with this approach is that the number of jobs (in quantum machine's queue) would increase in $O(n)=\Theta(n)$ complexity, where $n$ is the circuit depth, and the jobs themselves are small. Consequently, from a queuing latency viewpoint, this approach would not be preferable for users. However, from a reliability standpoint, it is the only method that actually {\em guarantees} reliable output (in the worst-case scenario with a slice size of one basic physical control-gate). For vendors, producing output under any input condition is probably the most desirable option, and they are responsible for the job policy, allowing them to make modifications to these policies while achieving the desired reliability.

In the DDC method on the other hand, the number of jobs is $o(klogn)=o(logn)$ (where $n$ is circuit depth and $k$ is a constant related to system reliability and the large order remains as $O(n)$), if the system is sufficiently reliable and the intended fidelity margin is large. Naturally, as we increase the fidelity threshold, we get closer to the output produced by the SDC, or we may not get any output at all (e.g., fidelity is 0.999 in a noisy NISQ). With a proper threshold (the minimum desired fidelity of a block) however, we would have larger blocks and fewer tasks, but with a risk to output reliability. While this approach would be more desirable from a user's perspective, there is an associated risk of 'incorrect output'  with the maximum count. Moreover, the circuit generation with this strategy requires more time compared to the SDC method, for implementing the DC. Based on i) the noise profiles of current systems, ii) the similarity levels of the simulation outputs to the actual system outputs, and iii) the time required to produce DC (via the DDC approach) due to the simulation hardware requirements and the latencies involved, {\em we currently prefer SDC over DDC}. However, the preference may change if/when the simulations become more accurate and less time/resource demanding in the future. 

%%%%% 

%% file: sections/6-evaluation.tex
%%%%%
\setlength{\tabcolsep}{2.0pt}
\begin{table*}[]
%\vspace{-2mm}
\centering  
\renewcommand{\arraystretch}{1.4}
\scriptsize

\begin{tabular}{l|lllll|lllll|lll|lll|}
\cline{2-17}
                                                  & \multicolumn{5}{c|}{\textbf{Baseline}}                                                                                                                                                  & \multicolumn{5}{c|}{\textbf{SDC}}                                                                                                                                                        & \multicolumn{3}{c|}{\textbf{DDC-60}}                                                                  & \multicolumn{3}{c|}{\textbf{SDC Worst Case}}                                                          \\ \cline{2-17} 
                                                  & \multicolumn{3}{c|}{\textbf{Classical Input}}                                                                                & \multicolumn{2}{c|}{\textbf{ghz state input}}               & \multicolumn{3}{c|}{\textbf{Classical Input}}                                                                                & \multicolumn{2}{c|}{\textbf{ghz state input}}               & \multicolumn{3}{c|}{\textbf{Classical Input}}                                                           & \multicolumn{3}{c|}{\textbf{Classical Input}}                                                           \\ \cline{2-17} 
                                                  & \multicolumn{1}{l|}{\textbf{PST {[}\%{]}}} & \multicolumn{1}{l|}{\textbf{E-Max(F)}} & \multicolumn{1}{l|}{\textbf{Job\#}} & \multicolumn{1}{l|}{\textbf{PST {[}\%{]}}} & \textbf{Job\#} & \multicolumn{1}{l|}{\textbf{PST {[}\%{]}}} & \multicolumn{1}{l|}{\textbf{E-Max(F)}} & \multicolumn{1}{l|}{\textbf{Job\#}} & \multicolumn{1}{l|}{\textbf{PST {[}\%{]}}} & \textbf{Job\#} & \multicolumn{1}{l|}{\textbf{PST {[}\%{]}}} & \multicolumn{1}{l|}{\textbf{E-Max(F)}} & \textbf{Job\#} & \multicolumn{1}{l|}{\textbf{PST {[}\%{]}}} & \multicolumn{1}{l|}{\textbf{E-Max(F)}} & \textbf{Job\#} \\ \hline
\multicolumn{1}{|l|}{\textbf{alu-v2\_30\_CXonly~\cite{Zhou_2020}}} & \multicolumn{1}{l|}{11.6\%}                & \multicolumn{1}{l|}{4.6\%}             & \multicolumn{1}{l|}{1}              & \multicolumn{1}{l|}{9.0\%}                    & 1             & \multicolumn{1}{l|}{85.1\%}                & \multicolumn{1}{l|}{77.7\%}            & \multicolumn{1}{l|}{45}             & \multicolumn{1}{l|}{85.3\%}                    & 91             & \multicolumn{1}{l|}{36.9\%}                & \multicolumn{1}{l|}{24.9\%}            & 9              & \multicolumn{1}{l|}{80.2\%}                & \multicolumn{1}{l|}{77.3\%}            & 45             \\ \hline
\multicolumn{1}{|l|}{\textbf{C17\_204\_CXonly~\cite{Zhou_2020}}}   & \multicolumn{1}{l|}{7.0\%}                 & \multicolumn{1}{l|}{3.4\%}             & \multicolumn{1}{l|}{1}              & \multicolumn{1}{l|}{3.4\%}                    & 1             & \multicolumn{1}{l|}{51.8\%}                & \multicolumn{1}{l|}{39.2\%}            & \multicolumn{1}{l|}{41}             & \multicolumn{1}{l|}{55.8\%}                    & 83             & \multicolumn{1}{l|}{44.6\%}                & \multicolumn{1}{l|}{37.9\%}            & 13             & \multicolumn{1}{l|}{59.1\%}                & \multicolumn{1}{l|}{52.2\%}            & 41             \\ \hline
\multicolumn{1}{|l|}{\textbf{cm82a\_208\_CXonly~\cite{Zhou_2020}}} & \multicolumn{1}{l|}{6.7\%}                 & \multicolumn{1}{l|}{2.0\%}             & \multicolumn{1}{l|}{1}              & \multicolumn{1}{l|}{4.7\%}                    & 1             & \multicolumn{1}{l|}{83.3\%}                & \multicolumn{1}{l|}{79.8\%}            & \multicolumn{1}{l|}{57}             & \multicolumn{1}{l|}{81.8\%}                    & 115             & \multicolumn{1}{l|}{57.4\%}                & \multicolumn{1}{l|}{51.5\%}            & 12             & \multicolumn{1}{l|}{80.7\%}                & \multicolumn{1}{l|}{78.0\%}            & 57             \\ \hline
\multicolumn{1}{|l|}{\textbf{ex2\_227\_CXonly~\cite{Zhou_2020}}}   & \multicolumn{1}{l|}{10.7\%}                & \multicolumn{1}{l|}{6.6\%}             & \multicolumn{1}{l|}{1}              & \multicolumn{1}{l|}{4.6\%}                    & 1             & \multicolumn{1}{l|}{71.5\%}                & \multicolumn{1}{l|}{64.7\%}            & \multicolumn{1}{l|}{55}             & \multicolumn{1}{l|}{69.7\%}                    & 111             & \multicolumn{1}{l|}{56.8\%}                & \multicolumn{1}{l|}{51.2\%}            & 11             & \multicolumn{1}{l|}{32.0\%}                & \multicolumn{1}{l|}{20.2\%}            & 55             \\ \hline
\multicolumn{1}{|l|}{\textbf{Multiplier-n-25}~\cite{li2020qasmbench}}    & \multicolumn{1}{l|}{2.8\%}                 & \multicolumn{1}{l|}{-8.3\%}            & \multicolumn{1}{l|}{1}              & \multicolumn{1}{l|}{9.1\%}                    & 1             & \multicolumn{1}{l|}{72.5\%}                & \multicolumn{1}{l|}{50.3\%}            & \multicolumn{1}{l|}{40}             & \multicolumn{1}{l|}{64.3\%}                    & 81             & \multicolumn{1}{l|}{64.7\%}                & \multicolumn{1}{l|}{42.3\%}            & 35             & \multicolumn{1}{l|}{0.0\%}                 & \multicolumn{1}{l|}{-72.8\%}           & 40             \\ \hline
\multicolumn{1}{|l|}{\textbf{qft\_16\_CXonly~\cite{Zhou_2020}}}    & \multicolumn{1}{l|}{1.0\%}                 & \multicolumn{1}{l|}{0.6\%}             & \multicolumn{1}{l|}{1}              & \multicolumn{1}{l|}{1.4\%}                    & 1             & \multicolumn{1}{l|}{68.3\%}                & \multicolumn{1}{l|}{59.8\%}            & \multicolumn{1}{l|}{48}             & \multicolumn{1}{l|}{58.2\%}                    & 97             & \multicolumn{1}{l|}{68.2\%}                & \multicolumn{1}{l|}{57.8\%}            & 10             & \multicolumn{1}{l|}{70.7\%}                & \multicolumn{1}{l|}{65.2\%}            & 48             \\ \hline
\multicolumn{1}{|l|}{\textbf{Average}}            & \multicolumn{1}{l|}{\textbf{6.6\%}}        & \multicolumn{1}{l|}{\textbf{1.5\%}}    & \multicolumn{1}{l|}{\textbf{1.0}}   & \multicolumn{1}{l|}{\textbf{5.4\%}}           & \textbf{1}    & \multicolumn{1}{l|}{\textbf{72.1\%}}       & \multicolumn{1}{l|}{\textbf{61.9\%}}   & \multicolumn{1}{l|}{\textbf{47.7}}  & \multicolumn{1}{l|}{\textbf{69.2\%}}           & \textbf{W7}    & \multicolumn{1}{l|}{\textbf{54.7\%}}       & \multicolumn{1}{l|}{\textbf{44.3\%}}   & \textbf{15.0}  & \multicolumn{1}{l|}{\textbf{53.8\%}}       & \multicolumn{1}{l|}{\textbf{36.7\%}}   & \textbf{47.7}  \\ \hline
\end{tabular}
\caption{Experimental results on the ibm\_cairo quantum machine~\cite{IBMQ}. }
\label{tab:my-table}
\vspace{-6mm}
\end{table*}

\section{Experimental Evaluation}
\label{sec:Evaluations}
This section evaluates DC on Ibm\_cairo a 27 qubits  Falcon\_r5.11 real quantum hardware~\cite{IBMQ}. For our evaluations, we choose circuits with sufficient depth to ensure that the original output reliability is sufficiently low. We then apply our DC (static Depth Control and dynamic Depth Control) to the target circuits, and report the output reliability result and the number of jobs, in each case. We present results for two types of inputs: classical state and superposition state. The classical state was the default benchmark circuit initialization -we did not touch the Benchmark. As for the superposition state, we opted for a GHZ state input, which is applied to \textbf{all} qubits. This entails utilizing a Hadamard gate followed by a sequence of CNOT gates (as shown in Figure~\ref{Fig:Superposition}), resulting in an expected input that equally likely generates either an all-zero or an all-one state.

\subsection{Methodology}
\label{subsec:Methodology}
The setup for our experimental evaluation is as follows. Our quantum classical circuits are obtained from the QASM benchmark suite~\cite{li2020qasmbench} and the Revlib implementation provided by Zhou et. al~\cite{Zhou_2020}. The Qiskit transpile's mapping and routing policies are set to 'SABRE' in all our experiments. Each target circuit in our experimental suite is evaluated on the ibm\_cairo~\cite{IBMQ} machine that has 27 physical qubits. Five thousand 'shots' are performed in each case. The $E-Max(F)$ metric represents the difference between the probability of obtaining the expected (correct) output and the probability of achieving the highest false output, expressed as a percentage. When used alongside PST, this metric helps to gauge the reliability of a system's output.

The 'worst-case' results for our SDC strategy are obtained using the Qiskit transpile 'basic' routing method and the 'trivial' layout-method. In this case, there are no routing and mapping optimization, and the mapping policy is 'direct map', e.g., it  maps logical qubit 0 to physical qubit 0; logical qubit 1 to physical qubit 1; etc. This may cause block's gates to map on non-neighboring qubits unnecessarily, which can be avoided by any optimization policy. In these experiments, the DDC fidelity threshold is set to 0.6, and the SDC block size is set to 5 gates. A block size of 5 gates was tested on a hypothetical worst-case design logic for the system, and it consistently produced reliable max counts under the 'sabre' routing and mapping policy. This range ensured reliable outcomes on the system, regardless of the circuit executed.

\subsection{Results}
\label{subsec:Result Discussion}
Table~\ref{tab:my-table} presents the PST and job counts for our two techniques, SDC and DDC, as well as the baselines. The results for single input show that SDC consistently provides the most reliable outcomes (higher than 50\% PST for the expected output) when the block size is kept small enough, as discussed in our theoretical analysis in Section~\ref{sec:Design}. In fact, With a block size of 5 (executing 5 gates per job), the system functions reliably, even in worst-case scenario of using direct mapping for the majority of cases. Our findings demonstrate up to $\sim{11x}$ PST improvement (reaching up to 85.1\% PST, on average 72.1\%) for single input with average $Avg(E-Max(F))=61.9\%$ providing reliable results when the baseline fails to deliver reliable output. For superposition state inputs, conclusions are consistent, showing an 11.4X improvement versus the baseline. With superposed inputs, initialization errors might alter DC's shot distribution, potentially impacting PSTs differently, though the average PST stays stable with reliable initialization. The PST calculation for superposed inputs, as per~\cite{Khadir2023TRIM}, is based on the average PST of expected outcomes. 

While SDC effectively improves PST, it does lead to a substantial number of jobs executed, resulting in extended queue times for the circuit. The number of jobs increases significantly for superposition-state input due to the greater quantity of executed circuits. Despite this, the number of shots executed in the superposition state is equal to that of the single-state input. As a result, we prefer to increase the block size and reduce job counts while at the same time minimizing error impact. It is crucial to note that, if the block size is not small enough, a false result could become dominant, leading to an error that propagates through to the final outcome. 

In the worst-case scenarios, the direct mapping policy is applied, and slicing is conducted {\em before} transpilation. The results with these scenarios are presented in Table~\ref{tab:my-table}. As demonstrated, Multiplier\-n25~\cite{li2020qasmbench} fails with no optimization on circuits as small as 5 gates; this failure occurred in one of the intermediate stages, so the error propagated to the final stage and led to an incorrect output with high probability. For example, when a CNOT is present between q0 and q26 in this scenario, the direct mapping positions them into q0 and q26 on the quantum machine, which are the farthest apart~\cite{IBMQ}, necessitating a considerable number of SWAPs for correct execution. As a result, even when slicing is done for 5 gates, a significant number of gates must be executed post-transpilation due to the added SWAPs. While other mapping and optimization techniques attempt to place them closer together, direct mapping processes the circuit without optimization to minimize transpilation time. Despite these worst-case conditions, the results still show that, in 5 out of 6 benchmarks, the correct outcome is reported with the highest PST. To address the error with no other optimization scenario, we recommend performing slicing after transpilation, ensuring that no significant number of gates are executed in a single job due to the added SWAPs. In post-transpile slicing, it is necessary to increase the block size to account for the additional SWAP operations, which will be included as part of the total gate count. 

Our results on single input state also show that Dynamic Depth Control (DDC) achieves a 54.7\% PST, while Static Depth Control (SDC) reported a higher PST of 72.1\%. DDC results indicate that it can produce reliable outputs $Avg(E-max(F)=44.3\%)$ when the threshold is adequately high and can concurrently reduce the number of jobs by approximately 3x. However, SDC demonstrates superiority over DDC due to the 'inconsistencies' between simulation and real hardware outcomes. For example, in the simulation of alu-v2\_CXonly~\cite{Zhou_2020}, the reported Probabilistic System Tolerance (PST) is over 60\%, but the actual hardware results show a PST of just 24\%, representing a 2.5x reduction in reliability. Given DDC's dependence on simulation, its results might be less dependable compared to SDC.

%%%%%

%% file: sections/7-conclusion.tex
%%%%%

\section{Conclusions and Future Works}
\label{sec:Conclusion}
In this study, we introduced DC, an approach that ensures reliable results for quantum classical circuits, regardless of their depth. DC  divides/slices a given quantum classical circuit into smaller, more reliable blocks, filters out noise, and confines the error rate of the quantum classical circuit to that of its final block. Additionally, DC offers super-linear speedups for mapping algorithms, primarily graph processing algorithms with exponential growth in size or state space, and creates opportunities for highly effective optimization algorithms by reducing the size of their input circuits. In addition, unlike the previously-proposed  quantum classical optimizations, our proposed modification makes DC~\cite{Paler2016Resizing,parent2015reversible,ding2020square} applicable to superimposed inputs. Our empirical data demonstrates an average improvement of approximately $\sim{11x}$ and $11x$ in PST for six benchmarks through slicing and noise mitigation on real quantum hardware, for classical input and GHZ-state input, respectively. Further, this is achieved while ensuring a significant contrast between the correct output and the second-most observed output, marking a $41x$ $E-Max(F)$ improvement.Therefore, we believe DC serves as an inspiration for future work, aiming to enable unlimited depth in NISQ systems when quantum memory becomes commercially available. Our planned  future studies include i) the manipulation of quantum memory technology as a noise filter to generalize our proposed approach to {\em all} quantum circuits and ii) the exploration of other potential technologies -- which can be more suitable -- for the development of quantum noise filters as independent blocks. 

%Circuit width depends on technology, specifically the number of qubits, while depth does not. Although both depth and size problems could theoretically be addressed through quantum memory, in its absence, approaches like~\cite{sadeghi2022quantum,mmmrOctober,hua2022exploiting} (with limitations on gate dependencies) are required for size/width optimization. To achieve infinite depth for all quantum circuits, we can employ quantum error correction (QEC) techniques along with quantum memory, creating analogs to DC to correct qubit states within reasonable intervals. As we currently only have access to classical memories, DC can be used to remove depth limitations for quantum classical circuits, as shown theoretically in Section~\ref{subsec:Heuristic} and experimentally in Section~\ref{sec:Evaluations}. 

%%%%%